\documentclass[doublecol]{epl2}
\usepackage{graphicx}% Include figure files

\title{Spin dynamics in triangular lattice antiferromagnets CuCr$_{1-x}$Mg$_x$O$_2$}
\shorttitle{Spin dynamics in triangular lattice antiferromagnets CuCr$_{1-x}$Mg$_x$O$_2$} %Insert here a short version of the title if it exceeds 70 characters

\author{Renwen Li\inst{1,2} \and Zhe Qu\inst{1}\thanks{To whom correspondence should be addressed; Email: \email{zhequ@hmfl.ac.cn}} \and Wei Tong\inst{1} \and Yuheng Zhang\inst{1,3}}
\shortauthor{Renwen Li \etal}

\institute{
  \inst{1} High Magnetic Field Laboratory, Chinese Academy of Sciences, Hefei 230031, China\\
  \inst{2} Department of Physics and Electronic Engineering, Hefei Normal University, Hefei 230061, China\\
  \inst{3} High Magnetic Field Laboratory, University of Science and Technology of China, Hefei 230026, China
}
\date{\today}
\abstract{
The electron spin resonance (ESR) spectroscopy was employed to investigate the spin dynamics in triangular lattice antiferromagnets CuCr$_{1-x}$Mg$_{x}$O$_2$ with $x =$ 0 and 0.02. All spectra can be well fitted by a single Lorentzian lineshape. The analysis of the $g$ factor, the linewidth $\bigtriangleup H$, and the ESR intensity $I$ as a function of temperature suggests the development of significant antiferromagnetic (AFM) spin fluctuations at temperature well above $T_N$ in both samples. However, the evolution of the AFM spin fluctuations is different for each sample. For undoped sample the ESR intensity $I$ is almost temperature independent between $\sim$ 100 K and 50 K and then drops rapidly below 50 K. But for $x =$ 0.02, the $I$ monotonously increases with cooling and reduces rapidly only below $T_N$. These results indicate that the AFM spin fluctuations are extremely strong in the undoped sample and appear to be suppressed upon Mg doping.}

\begin{document}

\maketitle

\section{Introduction}
%TLA attracts wide attention.ABO2 is a prototype system. CuCrO2 interesting.
Triangular lattice antiferromagnets (TLAs) have attracted considerable interests because geometrical spin frustration inherent to these systems results in exceptionally rich physical properties ranging over spin liquid, \cite{SrCrGaO,NiGa2S4} quantum phase transition, \cite{Cs2CuBr4} anomalous large thermoelectric response, \cite{NaCoO2TE} superconductivity, \cite{NaCoO2H2O} and magnetoelectric properties. \cite{CuFeO2ME,RbFe(MoO4)2ME,ACrO2ME}

\begin{figure}[t]
\includegraphics[angle=0,scale=1]{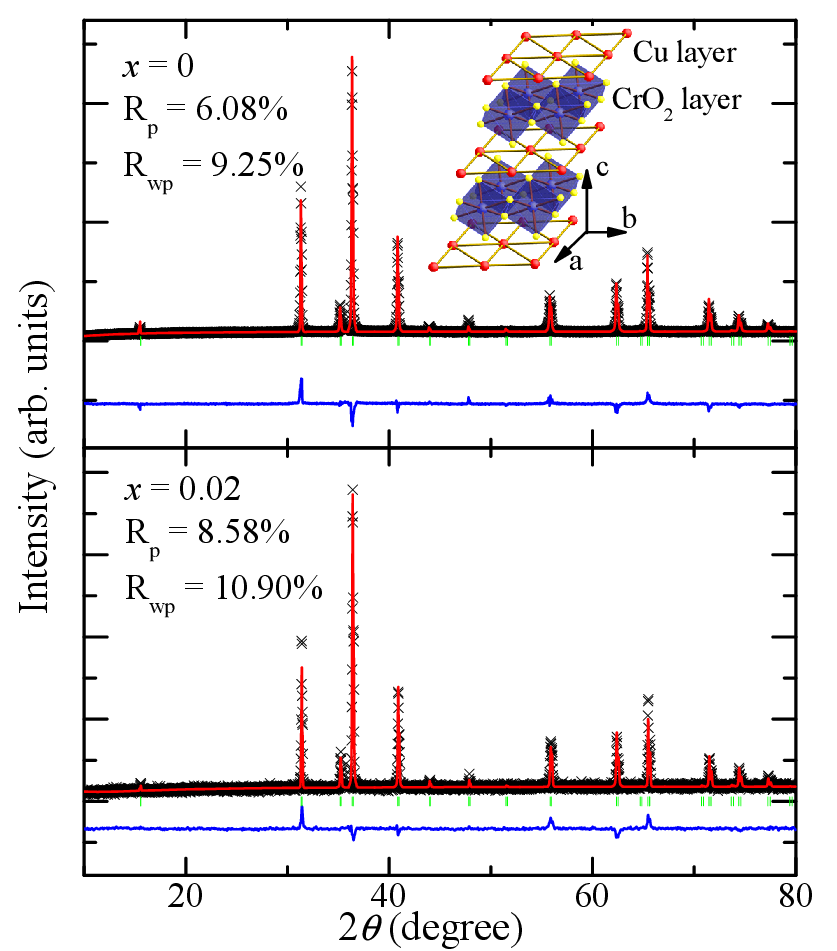}
\caption{(color online) Powder XRD patterns of CuCrO$_2$ and CuCr$_{0.98}$Mg$_{0.02}$O$_2$. The solid curve is the best fit from the Rietveld refinement using GSAS. The vertical marks indicate the position of Bragg peaks and the bottom curves show the difference between the observed and calculated intensities. Inset to the left panel shows the delafossite structure of CuCrO$_2$.}\label{fig:XRD}
\end{figure}

%CuCrO2
A typical example of TLAs is CuCrO$_{2}$; it has a delafossite structure, which can be viewed as the alternate stacking of edge-shared CrO$_6$ octahedral layers and Cu layers along the [001] axis (see the inset to Fig. \ref{fig:XRD}). Cr ions form an antiferromagnetic (AFM) triangular sublattice, and the compound enters a long range AFM ordered state below the N\'{e}el temperature $T_N \sim$ 26 K. \cite{CuCrO2TN} Previous powder neutron diffraction study proposed that the ground state has an out-of-plane incommensurate spiral-spin structure with an in-plane wave vector $q =$ (0.329, 0.329, 0) below $T_N$. \cite{CuCrO2neutron} In this spiral-spin ordered phase, the compound exhibits ferroelectricity driven by the magnetic order, \cite{ACrO2ME} whose polarization $P$ can be tuned by using both magnetic and electric fields. \cite{CuCrO2ME} In addition, for CuCrO$_2$ the temperature dependence of magnetization starts to deviate from the Curie-Weiss law at $\sim$ 200 K, around which a crossover occurs in the conductivity from a thermal activation behavior to variable range-hopping one. \cite{CuCrO2TN}

%doping
Magnesium doping appears to be quite unique in this system. Usually the introduction of non-magnetic dopants into a magnet will weaken the magnetic transition, $e.g.$ the Ca, Al doping in CuCrO$_2$. \cite{CuCrO2Cp,CuCrO2Cadoping} But the Mg substitution for Cr leads to the sharpen of the AFM transition and the large enhancement of magnetization around $T_N$, which means that the magnetic order is enhanced by Mg doping. \cite{CuCrO2TN,CuCrO2Cp} This is thought to stem from the partially lifted residual magnetic frustration due to spin fluctuations, which are enhanced through the interaction between the hole introduced by Mg doping and the localized Cr spins. \cite{CuCrO2TN,CuCrO2Cp}

%SRC important role. Previous neutron.
All these behaviors hint that there is a close relation between the magnetic and transport behaviors even at temperatures much higher than $T_N$ and the spin fluctuations might play an important role. Previous powder neutron diffraction probes observable intensity of (000) magnetic peak below $\sim$ 70 K, which is attributed to the occurrence of appreciable short-range magnetic interactions \cite{CuCrO2neutron}. However, the evolution of the spin fluctuations, and how the Mg doping will affect the spin fluctuations, are still not fully understood in CuCrO$_2$ system.

%ESR powerful tool. Here we report.
Electron spin resonance (ESR) spectroscopy has been known as a powerful probe for the study of static and dynamic magnetic correlations
on a microscopic level in strongly correlated materials. In this work, we use this tool to examine the micromagnestism in undoped and 2\% magnesium doped CuCrO$_2$. Our results suggest that the AFM spin fluctuations develop well above $T_N$ in CuCrO$_2$ but appear to be suppressed upon Mg doping.

\section{Experiment}

%Experiment
Polycrystalline samples of CuCr$_{1-x}$Mg$_{x}$O$_2$ with $x = $0 and 0.02 were prepared by the conventional solid-state reaction method described in \cite{ACrO2ME}. Stoichiometric proportions of high purity CuO (99.995\%), Cr$_2$O$_3$ (99.97\%) and Mg (99.9\%) were mixed and heated at 1000 $^{o}C$ in air with several intermediate grinding. Finally, they were pelletized, and then sintered at 1100 $^{o}C$ for 30 hours.
The structure and the phase purity of the samples were checked by powder X-ray diffraction (XRD) at room temperature. Magnetization measurements were performed with a commercial superconducting quantum interference device (SQUID) magnetometer (Quantum Design MPMS 7T-XL). The electron spin resonance spectra were collected using a Bruker EMXplus 10/12 CW spectrometer at X-band frequencies equipped with an Oxford continuous flow cryostat.

\section{Results and Discussion}

%XRD pure. DC magnetization consistent with previous results.
Fig. \ref{fig:XRD} displays the powder XRD patterns of CuCr$_{1-x}$Mg$_x$O$_2$ at room temperature. Rietveld refinement\cite{GSAS,EXPGUI} of the XRD patterns confirms that both samples are single phase with a hexagonal $R\overline{3}m$ structure. This refinement obtains lattice parameters, $a =$ 2.9751 ${\AA}$, $c =$ 17.1057 ${\AA}$ for $x =$ 0 and $a =$ 2.9765 ${\AA}$, $c =$ 17.1106 ${\AA}$ for $x =$ 0.02, which agree well with previous reports within the experimental error. \cite{CuCrO2TN}

\begin{figure}[t]
\centering
\includegraphics[angle=-90,scale=1]{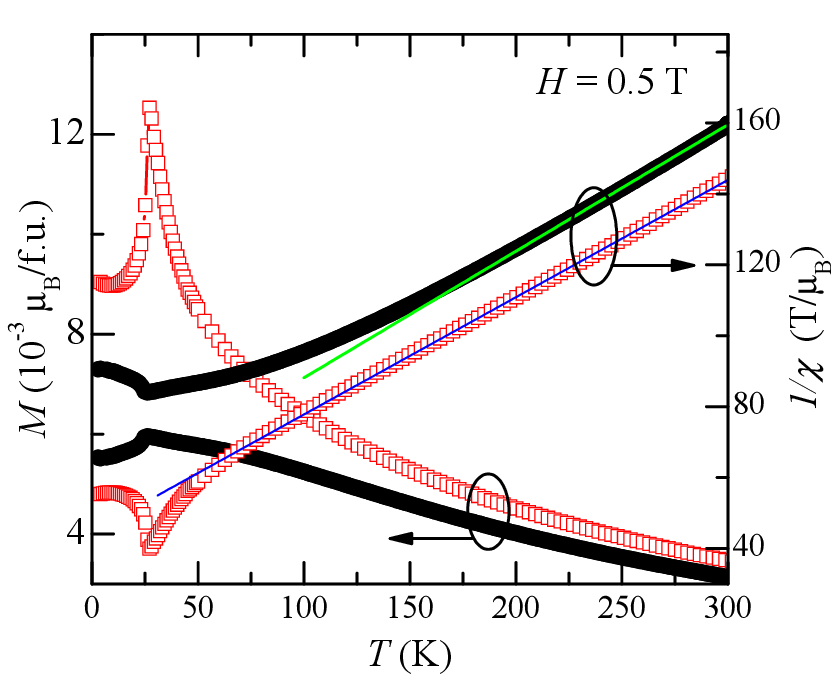}
\caption{(color online) Temperature dependence of the magnetization (M) and the reciprocal magnetic susceptibility (1/$\chi$)
of CuCr$_{1-x}$Mg$_{x}$O$_{2}$ with $x =$ 0 (closed circle $\bullet$) and 0.02 (open square %\textcolor{Red}{\square}\
) at 0.5 T, respectively. The straight lines are for the Curie-Weiss behavior.}\label{fig:MT}
\end{figure}

\begin{figure}[t]
\centering
\includegraphics[angle=-90,scale=1]{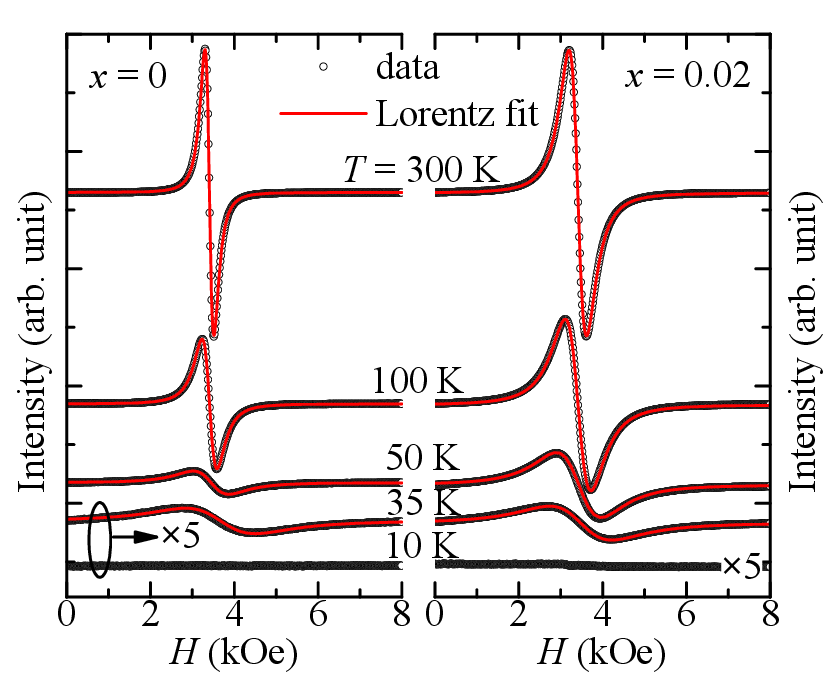}
\caption{(color online) Electron spin resonance (ESR) spectra of CuCr$_{1-x}$Mg$_x$O$_2$ shown at typical temperatures. The lines are displaced for clarification. Open circles $\circ$ represent the data and the solid lines show the fitting results to a Lorentzian lineshape. The intensity of the ESR spectra at 35 and 10 K for $x =$ 0 and at 10 K for $x =$ 0.02 are magnified by 5 times. }\label{fig:ESRspectra}
\end{figure}

The temperature dependence of the magnetization $M$($T$) is shown in Fig. \ref{fig:MT}. For the undoped sample, it reveals a characteristic downturn below 26 K, which is associated with the paramagnetic (PM) to AFM transition. Mg doping results in the sharpen of the AFM transition and the large enhancement of magnetization around $T_N$, suggesting the enhanced AFM magnetic order. We perform the Curie-Weiss analysis on these two samples, $i.e.$ $\chi \propto C/(T-T_{CW})$, where $C$ is the Curie constant and $T_{CW}$ is the Curie-Weiss temperature. The reciprocal susceptibility 1/$\chi$ is plotted as function of the temperature $T$, which is also shown in Fig. \ref{fig:MT}. It can be seen that for the undoped sample 1/$\chi$ shows an upward deviation of  from the linear below $\sim$ 200 K and the fitting yields a Curie-Weiss temperature $T_{CW} =$ -170 K, suggesting large AFM spin fluctuations. When Mg is introduced in the system, the temperature corresponding to the deviation of the $\chi^{-1}(T)$ from the linearity, and the Curie-Weiss temperature of $x =$ 0.02, change to 60 K and -135 K, respectively. All these results are consistent with previous reports, \cite{CuCrO2TN,CuCrO2Cp,CuCrO2ES} confirming that our samples are of high quality.

\begin{figure}[t]
\includegraphics[angle=0,scale=1]{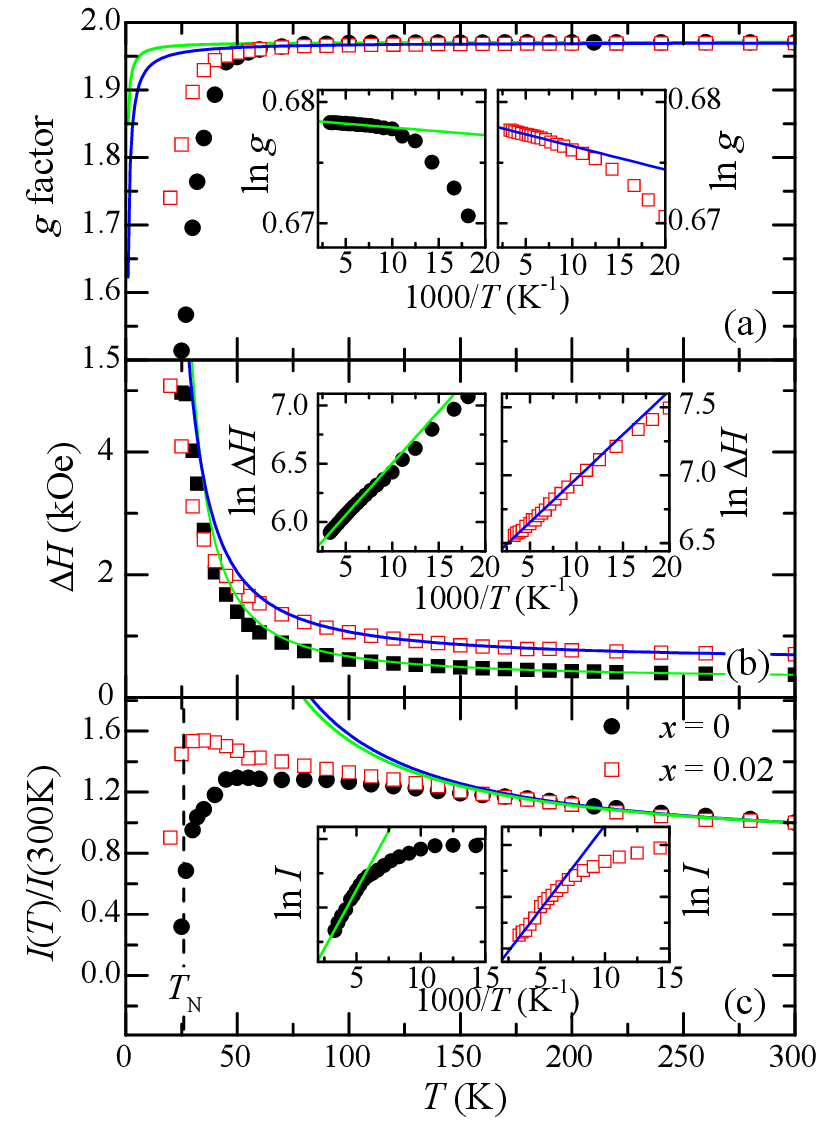}
\caption{(color online) The $g$ factor, the linewidth $\bigtriangleup H$, and the normalized ESR intensity $I$ by its value at 300 K, as function of temperature $T$ for CuCr$_{1-x}$Mg$_{x}$O$_{2}$ with $x =$ 0 and 0.02. The solid lines represent for the fitting results by the thermal activation model. Insets respectively display the $g(T)$, $\Delta H(T)$, and $I(T)$ replotted as $ln g$, $ln \Delta H$, and $ln I$ as function of 1000/$T$ for $x =$ 0 (upper inset) and 0.02 (lower inset).}\label{fig:ESRparameter}
\end{figure}

%ESR spectra and fitting process.
To obtain detailed information on the spin fluctuations, electron spin resonance spectra for CuCr$_{1-x}$Mg$_x$O$_2$ were measured. Fig. \ref{fig:ESRspectra} displays typical spectra at several temperatures. The spectra for all temperatures consist of a broad exchange-narrowed resonance line and can be very well fitted by a single Lorentzianline shape, indicating an insulating environment for Cr ions in CuCr$_{1-x}$Mg$_x$O$_2$. By performing such a fitting, the parameters including the $g$ factor, the linewidth $\bigtriangleup H$, and the ESR intensity $I$ can be obtained, which are summarized in Fig. \ref{fig:ESRparameter}.

%ESR parameters discussion

%g factor, linewidth
Let us first discuss the temperature dependence of the effective $g$ factor, which is shown in Fig. \ref{fig:ESRparameter} (a). At high temperature, the effective $g$ factor is close to 1.97, which is slightly below the free-electron value in accordance with the expectation for transition-metal ions with a less than half-filled $d$ shell. \cite{ESRgbook} With decreasing temperature, it monotonously decreases upon cooling. The deviation of the $g$ factor from the high temperature value can be described by a thermally activated model, $i.e.$, the Arrhenius law $\Delta g \propto exp(\Delta E/k_{B}T)$, where $\Delta E$ is the activation energy. An obvious deviation occurs below $\sim$ 100 K, which can be clearly seen in the insets to the Fig. \ref{fig:ESRparameter} (a). Such a deviation indicates that there is a rapid shift of the resonance field toward high field. The linewidth in CuCr$_{1-x}$Mg$_{x}$O$_{2}$ shows similar behavior to the $g$ factor (see Fig. \ref{fig:ESRparameter} (b)). For both samples, the linewidth increases with decreasing temperature and tends to diverge when the temperature approach $T_N$. The insets to Fig. \ref{fig:ESRparameter} (b) display the $\Delta H$($T$) replotted as $ln \Delta H$ $vs.$ 1000/$T$. It is found that at high temperatures the linewidth also obeys the Arrhenius law and again deviates from this relation below $\sim$ 100 K. Both the behaviors of the $g$ factor and the linewidth $\Delta H$ indicate the formation of a strong internal field above $T_N$ and suggests that the AFM spin fluctuations develop below $\sim$ 100 K.

%Intensity
The ESR intensity $I$ provides further information on the evolution of the AFM spin fluctuations. Fig. \ref{fig:ESRparameter} (c) displays the normalized ESR intensity $I$ by its value at 300 K as function of the temperature. It is well known that the ESR intensity in the PM regime is usually described by the Arrhenius law $I \propto exp(\Delta E/k_{B}T)$. \cite{ESRIArrhenius1,ESRIArrhenius2,ESRIArrhenius3} The Arrhenius plots of the intensity for both sample is shown in the insets to Fig. \ref{fig:ESRparameter} (c). For $x =$ 0, the ESR intensity first increases with decreasing temperature and can be fitted to an Arrhenius law at high temperatures (see the insets to Fig. \ref{fig:ESRparameter} (c)). But with further cooling, it deviates from the thermal activation behavior below $\sim$ 170 K and becomes almost temperature independent between $\sim$ 100 K and 50 K. Below $\sim$ 50 K, a fast suppression of the ESR intensity occurs. Since both the temperature independent behavior and the suppression of the ESR intensity occur at temperature much higher than $T_N =$ 26 K in the undoped sample, they can not be simply attributed to the establishment of the long-range AFM order. Instead, these intriguing phenomena should be related to the development of the AFM spin fluctuations at temperature much higher than $T_N$. As discussed before, the behaviors of the $g$ factor and the linewidth in CuCr$_{1-x}$Mg$_{x}$O$_{2}$ suggest the appearance of the AFM spin fluctuations below $\sim$ 100 K. The development of the AFM spin fluctuations will lead to the weakening of the ESR intensity, results in downward deviation from the Arrhenius law. With further cooling, the AFM spin fluctuations become increasingly stronger, making the ESR intensity almost temperature independent. When the temperature decreases below $\sim$ 50 K, the AFM spin fluctuations become extremely strong and eventually result in a rapidly suppression of the ESR intensity.

On the other hand, although the ESR intensity for $x =$ 0.02 also follows the thermal activation model at high temperatures and shows some downward deviation below $\sim$ 170 K, it monotonously enhances with decreasing temperature and only weakens rapidly when temperature decreases below $T_N$. Such a behavior is distinctly different from that for the undoped sample. One may concern whether such significant difference might be related to a possible magnetic impurity phase, MgCr$_2$O$_4$ spinel, which usually occurs in the Cu-Cr-Mg-O system \cite{CuCrO2TN,CuCrO2ES}. However, previous high field ESR results \cite{MgCr2O4ESR} indicate that the ESR intensity of MgCr$_2$O$_4$ decreases with decreasing temperature when the temperature decreases below $\sim$ 80 K, which clearly contradicts with the behavior observed in the Mg doped sample. Additionally, the characteristic peaks of MgCr$_2$O$_4$ are absent in the XRD pattern for CuCr$_{0.98}$Mg$_{0.02}$O$_2$ (see the lower panel in Fig. \ref{fig:XRD}). This indicates that even if the MgCr$_2$O$_4$ does exist the amount must be very tiny. The nominal composition of Mg could provide estimation for the upper limit of the amount of the impurity phase; $i.e.$ they could not exceed 2\% of the sample with all Mg forming MgCr$_2$O$_4$ spinel. Such tiny amount of MgCr$_2$O$_4$ could hardly introduce large impact on the results. Therefore, we can conclude that there is no observable contribution to the ESR from MgCr$_2$O$_4$ spinel in our sample and the significant change of ESR behavior induced by Mg doping must be intrinsic.

By comparison with the behavior of the ESR intensity for $x =$ 0, it is obvious that the ESR intensity for $x =$ 0.02 increases faster below $\sim$ 100 K, which suggests that although the AFM spin fluctuations also appears in $x =$ 0.02 at temperatures well above $T_N$, they are much weaker than that in $x =$ 0. This result means that magnesium substitution for Cr appears to suppress the AFM spin fluctuations in the system. Such an evolution of spin fluctuations with Mg doping is in contradiction with previous expectation that spin fluctuations will be enhanced upon Mg doping \cite{CuCrO2TN,CuCrO2Cp} Therefore, the mechanism behind the intriguing magnetic and transport properties in this system need to be reconsidered by taking into account the experimental results discussed above.

\section{Conclusion}
In summary, we have performed a comprehensive ESR study to investigate the spin dynamics in CuCr$_{1-x}$Mg$_{x}$O$_{2}$. The analysis of the $g$ factor, the linewidth $\Delta H$ and the ESR intensity $I$ as function of the temperature suggest that significant AFM spin fluctuations appear at temperature well above $T_N$ in both samples. The evolution of the spin fluctuations with temperature and Mg doping is mapped out. Our results indicate that the AFM spin fluctuations are extremely strong in the undoped sample and appear to be suppressed upon Mg doping.

\acknowledgments
This work is supported by the State Key Project of Fundamental Research of China under contract No. 2007CB925001 and 2010CB923403.

\end{document}